\documentstyle[prb,aps,multicol,epsf]{revtex}
\begin{document}

\draft
\title{Internal Friction of Amorphous Silicon in a Magnetic Field}

\author{T. H. Metcalf, 
Xiao Liu\cite{authorLiu}, and R. O. Pohl\cite{authorPohl}}
\address{Laboratory of Atomic and Solid State Physics, Cornell 
University, Ithaca, New York 14853}
\date{\today}
\maketitle
\begin{abstract}
The internal friction  of e-beam amorphous silicon was 
measured in a magnetic field between 0 and 6 T, from  1.5--20 K, and
 was found to be independent of the  field to better 
than 8\%. It is concluded that the low energy excitations observed in 
this experiment are predominantly atomic in nature.
\end{abstract}
\pacs{71.23.Cq}

\begin{multicols} {2}
\narrowtext
It is well known that amorphous solids have a broad spectrum of 
low-energy excitations. At low temperatures, characteristic 
signatures of these excitations can be seen in a variety of thermal 
(e.g. specific heat, thermal conductivity) and elastic (e.g. internal 
friction, ultrasound attenuation, sound velocity) 
measurements\cite{arup}. They are  commonly described with the two-level 
tunneling  model\cite{phillips,anderson,jaeckle}, in which
it is assumed that
the disordered structure of the 
material permits atoms or groups of atoms to tunnel between two 
spatial
positions in close proximity and with energy splittings spanning a 
wide temperature range. The 
microscopic nature of these tunneling states, however, is still unknown.

It has been suggested that tetrahedrally 
bonded amorphous solids ({\it a}-Ge and {\it a}-Si) are structurally
overconstrained and will therefore 
have a lower density of tunneling states, or perhaps none at 
all\cite{phillips}. 
Experimental searches for evidence of tunneling states (or lack 
thereof) in such materials has been difficult, as these materials can 
only be grown as thin films on substrates. In practice, most 
experimental techniques employed must compare measurements of the 
bare substrate (e.g. its heat capacity in the case of specific heat) 
to those of the substrate plus the thin film. The value for 
the film itself is then extracted by considering the film-substrate 
geometry. Since the addition of the film to the substrate will often
produce only a small change in the raw measurements, the sensitivity 
of these measurements is necessarily limited.
Indeed, experiments trying to determine the existence of such states 
have  given inconsistent results: some experiments showed 
evidence for such states, and some did not, as
recently reviewed\cite{LiuAndPohl}.

Of particular relevance to the present work were low-temperature specific
heat measurements with e-beam  {\it a}-Ge  by van den 
Berg and v.~L\"ohneysen\cite{l''ohneysen}. 
Specific heat is  a direct measure of the excitations 
that exist within a solid---the two level 
tunneling model predicts a contribution 
  linear in temperature which at sufficiently low temperatures 
will dominate the $T^{3}$ phonon contribution.  However, the presence of 
specific heat  in excess of the phonon contribution is 
not necessarily the result of tunneling states.  One early 
example\cite{zeller} was the specific heat of several silica-based glasses, 
which showed excitations in addition to two-level tunneling states.   These
additional excitations vanished in the presence 
of a moderate magnetic field, and were attributed to spins from 
iron impurities\cite{stephens}.
Although {\it a}-Ge was
indeed shown to 
have an excess specific heat below 1 K, it almost completely vanished
in the presence of 
a 6 T magnetic field\cite{l''ohneysen}. Hence the 
extra excitations were concluded to be electronic---not atomic---in 
nature, and were attributed to exchange coupled clusters of dangling 
bonds, which experienced Zeeman splitting in a magnetic field. However, 
for the reasons mentioned above, these experiments were not sensitive enough
to 
completely rule out a separate magnetic-field-insensitive linear
contribution underlying the 
electronic contribution to the specific heat, but could only be used 
to determine an upper bound to its magnitude.

Liu and Pohl\cite{LiuAndPohl}, of whose work this paper is an 
extension, probed the existence of low energy excitations in {\it a}-Si and 
{\it a}-Ge films through measurements of internal friction at low 
temperatures using silicon double-paddle oscillators as substrates. A 
bare double-paddle oscillator itself has an extremely small internal 
friction background, typically $Q^{-1}=2\times10^{-8}$ at liquid 
helium temperatures\cite{LiuAndPohl}. The internal friction of a thin 
film will then 
dominate---and not be a small perturbation to---the total damping of 
a film-carrying paddle. Generally, amorphous solids have a 
temperature-independent internal friction ``plateau,'' whose width 
extends from 
approximately 100 mK to 10K, depending on the measuring 
frequency\cite{ToppAndCahill}. The prototypical amorphous solid, 
{\it a}-SiO${}_{2}$, has an internal friction $Q^{-1}=4\times10^{-4}$ 
in this plateau, and almost all amorphous 
solids have internal frictions within a factor of 3 of this value 
(referred to as the ``glassy range'').
For e-beam {\it a}-Ge films, $Q^{-1}=0.7\times10^{-4}$, close to 
the glassy range\cite{LiuAndPohl}. Based on this value and using the 
tunneling model, the specific heat   was calculated
to be close to the upper limit determined 
experimentally in  large magnetic 
fields\cite{l''ohneysen}. However, since the internal friction had 
been measured in zero $B$-field\cite{LiuAndPohl}, one has to ask whether
the states seen were truly atomic in nature and not electronic.

To this end, we measured the internal friction of paddle oscillators 
with e-beam {\it a}-Si films in a ${}^{4}$He cryostat above 1.4 K and in magnetic fields as large as 6 T. In a magnetic 
field $B$, an electronic excitation  
will have  a Zeeman splitting $g\mu_{B}B$ raising its excitation temperature
by $g\mu_{B}B/k_{B}$. Using $g=2$ and $B=6\:\rm T$,  this temperature 
is approximately 8 K. Therefore, any damping resulting
\begin{figure}[t]
\begin{center}\leavevmode
\epsfxsize=3 in
\epsfbox{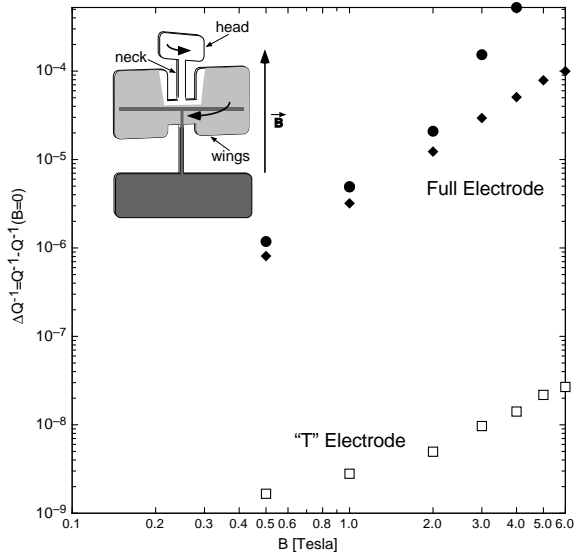}
\vspace{1 cm}
\caption{Enhancement of background internal friction due to magnetic 
field in three paddles without {\it a-}Si films. Solid diamonds: full 
electrode coverage, 2.4 K; solid circles: full electrode coverage, 
6.8 K; open squares: `T' electrodes, 4.9 K. Zero field background 
internal friction is $2\times10^{-8}$ for full electrode paddles and 
$1\times10^{-8}$ for the `T' 
electrode paddle is  (see text for details).
The inset, a drawing of the paddle, shows full electrode 
coverage (shaded areas) and `T' electrode coverage (darker shaded 
areas). Arrows on head and wings describe the antisymmetric mode studied 
in this work.
B-field direction is parallel to the neck.}
\label{figure1}\end{center}\end{figure}
\noindent
from the relaxation of electronic 
excitations---most likely to occur by 
one-phonon emission---should be drastically reduced 
at sufficiently low temperatures. 

Details of 
sample preparation and measurement are the same as described 
previously\cite{LiuAndPohl}. 
The following changes were made to the apparatus for work in a 
magnetic field, which was applied parallel to the paddle's axis of 
rotation. 
First, we replaced the ferromagnetic Invar block to which the 
paddle is usually attached with one made of 
silicon in order to avoid disturbance of the field.
Second, we found that the usual metal layer deposited on the back of 
the paddle and which acts as an electrode---30 ${\rm\AA}$ Cr followed by 
500 ${\rm\AA}$ 
Au covering an area indicated in gray shading in the inset to 
Fig. 1---caused an unacceptably large background damping in 
a magnetic field. According to Ref. \onlinecite{LiuAndPohl}, the 
presence of a $0.7\:\rm\mu m$ {\it a-}Si film 
is estimated to increase the zero-field internal friction to 
$Q^{-1}_{\text{paddle}}\approx4\times10^{-7}$.
At a field of only 1 T, however, the bare 
paddle background internal friction was 
$Q^{-1}=5\times10^{-6}$, over an order of
magnitude larger than 
that of the film-carrying paddle in zero field. 
The magnetic field 
dependence of this background internal friction
is shown in Fig. 1 for two different paddles at constant temperatures.
No attempts were 
made to 
quantatively account for the large magnitude of this damping, which 
\begin{figure}[t]
\begin{center}\leavevmode
\epsfxsize=3 in
\epsfbox{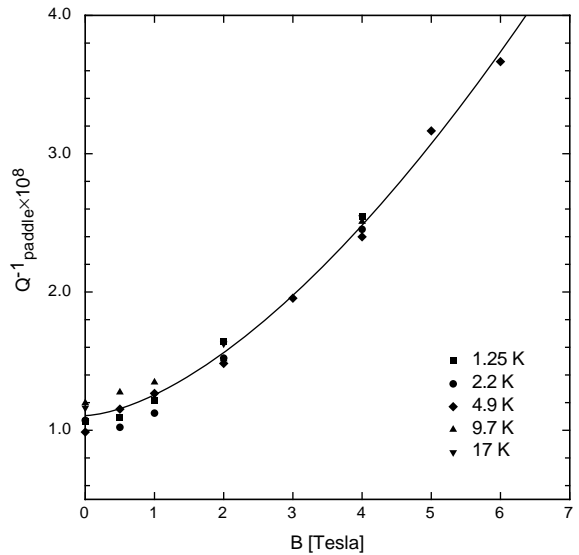}
\vspace{.5cm}
\caption{Background internal friction  {\it vs.}\ magnetic field for 
`T'-electrode paddle for several temperatures. Also shown is fit from 
eq. (\protect\ref{dampingEquation}.)}
\label{figure2}\end{center}\end{figure}
\noindent
is likely due to eddy currents. However, by 
making the metal film electrode in the shape of a 
thin `T,' shown in dark gray shading in the inset to Fig. 1, 
the large field effect was reduced by three orders of magnitude, 
with $Q^{-1}_{\text{sub}}$ rising only by a factor of 4 in a field 
of 6 T. Furthermore, this 
background internal friction was temperature-independent between 1 
and 20 K, as is 
shown in Fig. 2, and could be described simply
as the following function of magnetic field:
\begin{equation}
Q^{-1}= 1.106 \times 10^{-8} + 1.503 \times 10^{-9} \, 
B^{1.60}\label{dampingEquation},
\end{equation}
where the magnetic field $B$ is measured in Tesla. Note that even for 
zero magnetic field, the background internal friction is decreased by a 
factor of 2 below that observed with the large metal 
electrode\cite{LiuAndPohl}. Since metal films are known to have large 
internal frictions\cite{liumetals}, the reduction in the size of the metal film 
electrode can explain the observed reduction of the damping.

Since the paddle oscillators themselves are very fragile, the 
preparation must follow a certain order: the {\it a}-Si film must be 
deposited before epoxying the oscillator to the silicon block, and the metal 
electrodes deposited thereafter. Once affixed to the block, a paddle
cannot be removed without a large risk of breaking it. 
Hence we had to use two different oscillators for the
measurements with and without films. 
Considering the strong dependence of the background 
damping on details of electrode shape and the
irreproducibility in depositing the `T' electrodes, the background 
damping of the two oscillators used might differ. After comparing two 
different bare 
oscillators with `T' electrodes, we concluded that such differences are  
no larger than 25\% over the entire range
\
\begin{figure}[t]
\begin{center}\leavevmode
\epsfxsize=3 in
\epsfbox{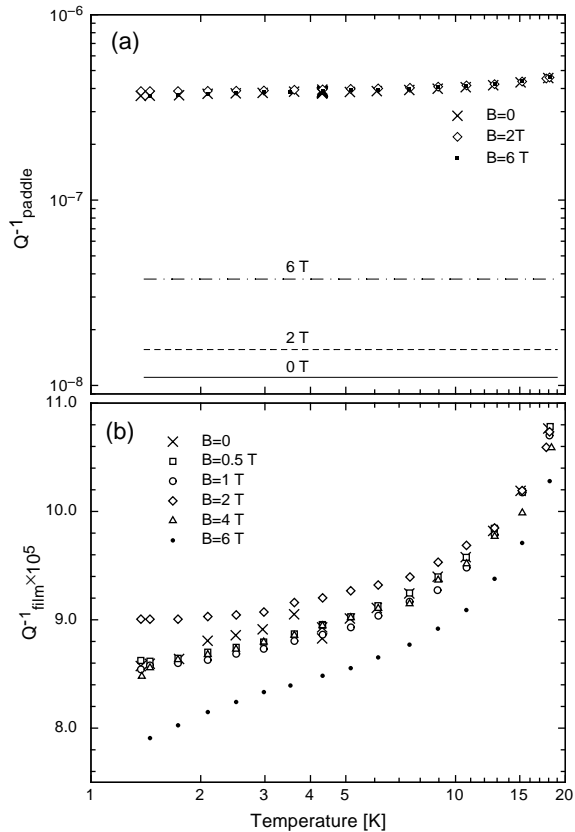}
\vspace{.5cm}
\caption{(a) Internal friction of film-carrying paddle as a function of 
temperature 
in 0, 2, and 6 T, together with temperature-independent background  computed from 
eq. (\ref{dampingEquation}). (b) Temperature dependence of film 
internal friction, with
background subtracted via eq. (\ref{filmEquation}).}
\label{figure3}\end{center}\end{figure}
\noindent
of magnetic field
and thus should not significantly affect the analysis.

We extract the internal friction 
of the film itself using\cite{LiuAndPohl}
\begin{equation}
Q^{-1}_{\rm film}= {G_{\rm sub}t_{\rm sub}\over 3 
G_{\rm film}t_{\rm film}}\left(Q^{-1}_{\rm paddle}-Q^{-1}_{\rm 
sub}\right)\label{filmEquation}.
\end{equation}
The 
thickness of the paddle oscillator substrate is $t_{\text{sub}}=300\,\mu\rm m$, 
the 
shear modulus of crystalline silicon in the direction of the neck, 
(110), is taken as 
$G_{\rm Si}=6.2\times 10^{11}{\:\rm dynes/cm^{2}}$, and that of the 
film 
$G_{\rm film}=3.63\times 10^{11}\:\rm dynes/cm^{2}$ (Ref. 
\onlinecite{LiuAndPohl}).
Film thickness was $t_{\rm film}=708\:\rm nm$ in this measurement.

The internal friction of the film-carrying paddle for B= 0, 2, 
and 6 T is plotted in Fig. 3a, together with the temperature
independent background 
internal friction  given by eq. (\ref{dampingEquation}). In the 
absence of a magnetic field, the 
addition of the film 
increases the total damping by over one order of magnitude, from 
$ Q_{\rm paddle}^{-1}=1\times10^{-8}$ to $Q_{\rm 
paddle}^{-1}=4\times10^{-7}$.
In a magnetic field, the damping is nearly unchanged, which indicates that 
most of the damping is not caused by electrons, but by atomic
relaxations.
\begin{figure}[t]
\begin{center}\leavevmode
\epsfxsize=3 in
\epsfbox{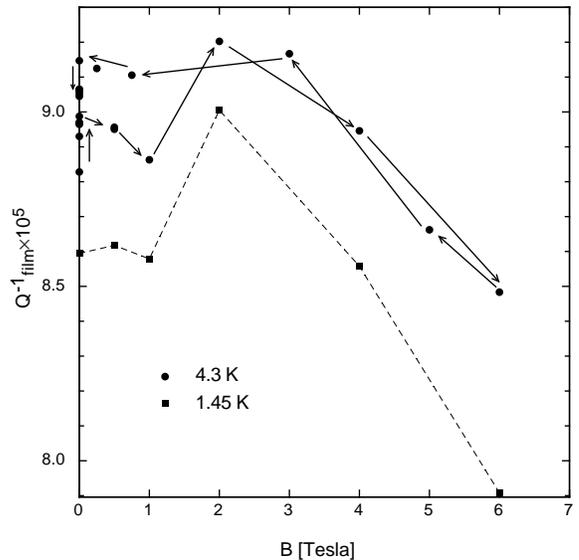}
\caption{Magnetic field dependence of the internal friction of the 
{\it a-}Si film at 4.3 K and 1.45 
K. Time history of the 4.3 K data is traced.}
\label{figure4}\end{center}\end{figure}
\noindent 
For a quantitative estimate of the upper limit of the effect of electrons, 
the internal friction of the film itself is computed using eq. 
(\ref{filmEquation}). In zero field, $Q^{-1}_{\rm 
film}=8.5\times10^{-5}$ at 1.5 K, increasing with increasing 
temperature to $Q^{-1}_{\rm film}=1.1\times10^{-4}$ at 18K, as shown 
in Fig. 3b. These 
values are 30\% smaller than those reported in
Ref. \onlinecite{LiuAndPohl}, but 
this can be explained considering the large sensitivity of the 
internal friction to even moderate heat treatment noted previously in 
these films. 
Although the film-carrying paddle 
shows virtually no variation with magnetic field, the fact that 
$Q^{-1}_{\rm sub}$ increases with magnetic field  implies a 
small decrease of $Q^{-1}_{\rm film}$ with increasing magnetic field.
The variation of $Q_{\rm film}^{-1}$ with increasing 
$B$, however, is not simple; it is shown in more detail for two temperatures 
in Fig. 4. In both cases, $Q^{-1}_{\rm film}$ has a maximum near 2 T, 
followed by a decrease with larger fields. The data were taken over 
the course of several days, and for 4.3 K, the time history of the 
data is traced. Some hysteresis is observed; the initial 
zero-field data have lower damping than the final zero-field data by 
approximately 4\%. Whether this hysteresis and the apparent field 
effect are real or artifacts of the background measurement or other 
experimental details cannot presently be decided. 

The very small  changes observed in our experiments  
are to be 
compared with the specific heat measurements.  Writing their linear 
specific heat term as $C=aT$, van den Berg and 
v.~L\"ohneysen\cite{l''ohneysen} found in zero field
$a=4.5\times10^{-6}\rm\, J/g\,K$, while in 6 T, 
$a\leq1\times10^{-7}\,\rm J/g\,K$, {\it i.e.}\ a suppression by at least 
97\%. Similarly, Stephens\cite{stephens} observed a complete removal 
of the specific heat anomaly caused by 12 ppm of iron in a 
borosilicate glass below 2K in a magnetic field of $B=3.3\rm\: T$. 
Biggar and Parpia studied the magnetic field dependence of the 
internal friction of boron-doped crystalline silicon\cite{biggar}, in which
  phonon 
scattering from holes bound to acceptor (boron) atoms causes a large 
internal friction in the absence of a magnetic field. This excess 
damping was observed in 
paddle oscillators fabricated from such crystals, and was reduced by 98\%
in a 6 T magnetic field.
Compared with these 
experiments, the suppression observed in our experiments is 
insignificant. 
It is therefore concluded that electrons play only a negligible role 
in our experiments and that the relaxation observed is caused by 
atomic motion as predicted by the tunneling model.

We would like to thank Jeevak Parpia for use of the cryostat in which 
these experiments were carried out, and Rob Biggar for help with the 
experiment. We also thank them and Richard Crandall for stimulating 
discussions. This work was supported by the National Science 
Foundation Grant No. DMR 9701972, and by the National Renewable Energy 
Laboratory, Grant No AAD-9-18668-12.

\end{multicols}

\end{document}